\documentstyle[epsfig]{europhys}

\newif\ifboo \boofalse

\def\Review#1{\boofalse{\it #1},}

\def\Vol#1{\ifboo Vol. {\bf #1}\else{\bf #1}\fi}
\def\Year#1{\ifboo #1\else(#1)\fi}

\def\Page#1{\ifboo {\rm p. #1}\else{\rm #1}\fi}

\begin{document}

\euro{}{}{}{}

\Date{}

\title{The Many Electron Ground State of the Adiabatic \\
Holstein Model in Two and Three Dimensions}

\author{B. Poornachandra Sekhar, Sanjeev Kumar and Pinaki Majumdar}

\institute{ Harish-Chandra  Research Institute,\\
 Chhatnag Road, Jhusi, Allahabad 211 019, India }

\rec{}{}


\maketitle

\begin{abstract}

We present the complete ground state phase diagram of the Holstein model in
two and three dimension considering the phonon variables to be classical. We 
first establish the overall structure of the phase diagram by using exact 
diagonalisation based Monte Carlo (ED-MC) on small lattices and then use a 
new ``travelling cluster'' approximation (TCA) for annealing the phonon 
degrees of freedom on large lattices. The phases that emerge include a Fermi 
liquid (FL), with no lattice distortions, an insulating polaron liquid (PL)  
at strong coupling, and a charge ordered insulating (COI) phase around 
half-filling. The COI phase is separated from the Fermi liquid by a regime of 
phase coexistence whose width grows with increasing electron-phonon coupling. 
We provide results on the electronic density of states, the COI order 
parameter, and the spatial organisation of polaronic states,  for arbitrary 
density and electron-phonon coupling.  The results highlight the crucial role 
of spatial correlations in this strong coupling problem.

\end{abstract}

\section{Introduction}
Electron-phonon (EP)  interactions are ubiquitious in metals and
dominate the finite temperature resistivity in most electron
systems. While the perturbative, Fermi liquid, regime in
EP systems is well understood \cite{ziman}, the electronic
ground state is fundamentally reorganised when EP
interactions are strong. 
The presence of a strong local interaction can generate a large
lattice distortion, creating a potential well for
the electron, and lead to a `self trapped' polaronic state 
\cite{old-pol}.
Strong electron-phonon coupling, sometimes in conjunction with
other interactions,  form a crucial component in
the physics of several correlated systems, {\it e.g}, the manganites
\cite{mang-ref},
the nickelates \cite{nick-ref}, and the traditional charge density wave systems
\cite{cdw-ref}.

The full `polaron problem', considering the quantum dynamics of the
phonons and strong EP coupling, 
is well understood 
for a single (or few) electrons 
\cite{pol-ref1,pol-ref2,pol-ref3,pol-ref4,pol-ref5}.
The non perturbative nature of the problem and the exponential growth
in Hilbert space with electron number, $N_{el}$, has made the finite
density problem difficult to access.  
Usually, the physical interest is in the 
 ``adiabatic'' regime 
since typical phonon 
frequencies in most materials are  much smaller than 
the electron hopping scale.
Although adiabaticity by itself does not lead to a simpler problem, 
the {\it adiabatic limit}, where phonon dynamics 
is ignored, leads to a relatively tractable situation. 
This limit, of electrons coupled to classical phonons, 
has been explored within dynamical
mean field theory (DMFT) \cite{millis-mull,ciuchi-cdw,millis-cdw}. 
DMFT,  however, 
loses out on spatial correlations or the possibility of accessing 
non periodic 
phases, important
at strong coupling.  In this
paper we use an approach that 
{\it explicitly retains spatial correlations }
and set out the ground state  
of many electron systems, in two and three dimension, coupled to
adiabatic  phonons. 

Let us define the Holstein model, whose adiabatic  limit we explore:
\begin{equation}
H = -t\sum_{\langle ij \rangle}  
c^{\dagger}_i c^{~}_j 
-  \mu \sum_{i }  n_i  
- \lambda \sum_i n_i  x_i 
+ {1 \over {2M}} \sum_i p_i^2 + {K \over 2} \sum_i x_i^2
\end{equation}
The $t$ are
nearest neighbour hopping on a $d$ dimensional lattice,  
$\mu$ is the chemical potential, and   
$n_i = c^{\dagger}_i c_i$ is the electron density operator
for spinless fermions.
The phonon coordinate is $x_i$, and  $p_i$ is 
the momentum conjugate to $x_i$, 
with $[x_i , p_j]=i \delta_{ij}$. $M$ and $K$ are respectively the
`mass' and stiffness of the phononic oscillators, and $\lambda$ is the
EP coupling.
We set $t=1$, as our reference scale, put $K=1$  and also $\hbar =1$.
The adiabatic limit sets $M \rightarrow \infty$.
For ``quantum''  phonons,  with
$\omega_{ph} = \sqrt{K/M}$, the last three terms in $H$
can be written as: 
$- \lambda_Q \sum_i n_i (b^{\dagger}_i + b_i)  
+ \omega_{ph} \sum_i b^{\dagger}_i b_i $, using  
$x_i = (b_i
+ b^{\dagger}_i )/\sqrt{2M \omega_{ph}}$,
and  $\lambda_Q = \lambda
\sqrt{{\omega_{ph}}/(2K)}$. 

The  parameter space of the model is defined by: 
$(i)$~the `coupling parameter' 
$\alpha= E_p/2dt$, with $E_p= \lambda^2/2K$,
$(ii)$~the `adiabaticity' $\gamma= \omega_{ph}/t$,
 and $(iii)$~electron density, $n$, in addition to 
dimensionality, $d$.
The parameter $\alpha$ 
quantifies the competetion between trapping and delocalisation,
while $\gamma$
measures  the `slowness' of the phonons compared to  electron
dynamics.  

The  ground state 
of this model is understood only
in a few limiting cases:  
$(i)$~At weak to moderate coupling, traditional
Migdal-Eliashberg (ME) theory \cite{migd-el}
 can be used for the many electron problem for $\gamma \ll 1$,
and describes a Fermi liquid (FL) metal (or a superconductor,
when electron spin is also considered).
$(ii)$~At strong coupling, and for arbitrary $\gamma$, there is
no controlled analytic theory even for a single electron
coupled to the phonon system. There are, however, 
powerful numerical methods
and the `single polaron' ground state 
is essentially understood 
\cite{pol-ref1,pol-ref2,pol-ref3,pol-ref4,pol-ref5}.
Little is known of the many electron problem beyond ME theory.
$(iii)$~Controlled theory of strong coupling finite 
density systems  
has been possible only in the classical limit, $\gamma =0$,
in the limit $d \rightarrow \infty$. The original study by 
Millis {\it et al.}  \cite{millis-mull}
highlighted the FL and polaronic insulator phases,  
within 
DMFT, and other DMFT studies have  explored  
charge ordering \cite{ciuchi-cdw,millis-cdw}. 

While contributing to the initial understanding,  DMFT misses
out on some crucial aspects of the physics.
$(i)$~The
spatial correlation of lattice distortions, and the 
resulting correlations between polarons, when they form,
is not accessible, 
and
$(ii)$~the possibility of phase coexistence and cluster formation
is hard to access.
We need to move beyond DMFT to recover these physical effects, and
address current materials issues.
The method we use, described next, 
can handle  
many electron systems strongly coupled to adiabatic phonons,
at arbitrary  disorder, explicitly in $1-3$ dimension.
This paper focuses on the ground
state of `clean' systems, 
subsequent papers discuss the finite temperature effects  and
the role of disorder \cite{sk-pm-latt-pol2}.
\begin{figure}

\hspace{1.5cm}
{\epsfxsize=11.0cm, \epsfysize= 5.8cm, \epsfbox{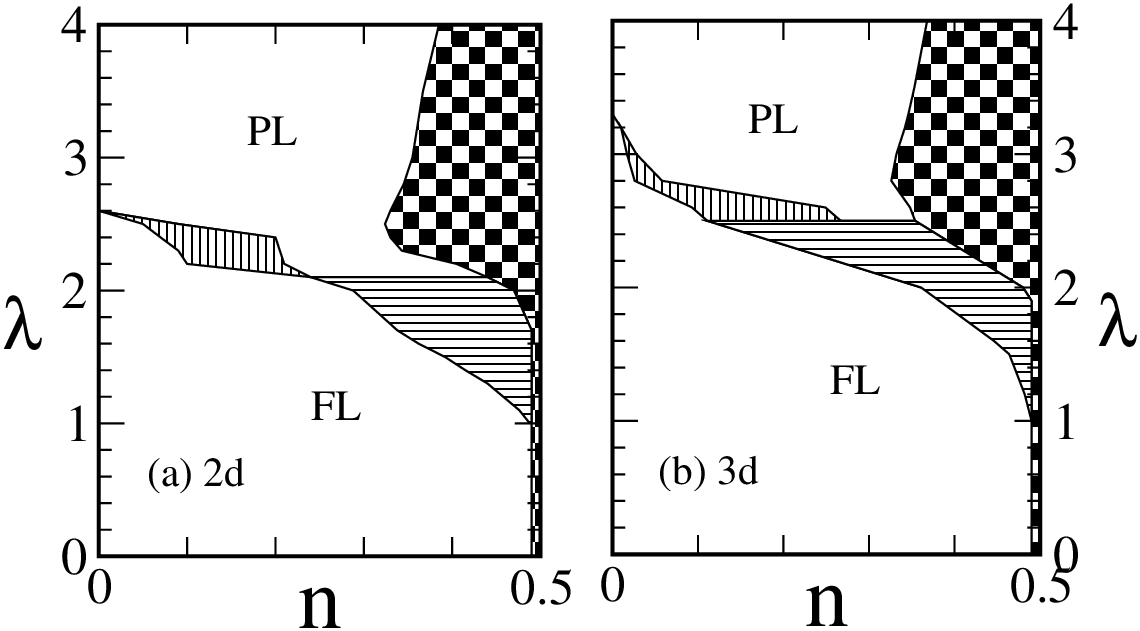}}

\caption{Phase diagrams at $T=0$. FL indicates Fermi liquid (band metal), the
chessboard pattern represents a charge ordered insulator 
with  $\{ \pi, \pi, ..\}$ modulation, and PL is a 
positionally disordered insulating polaron liquid.
Shading indicates regimes of phase separation.
The phase diagram are obtained 
through a combination of TCA and analytic estimates: 
$(a)$ result on 2d, TCA with $L_c = 4$ and system size $24^2$, 
$(b)$~result on 3d with $L_c = 4$ and system size  $ 8^3$.
The coexistence width at weak coupling is  estimated as explained
in the text. }

\vspace{-2mm}

\end{figure}

\section{Method}
At low temperature the adiabatic 
EP  problem reduces to finding the phonon
configuration that would minimise the total energy:
$
{\cal E}\{x\} = 
\langle 
-t \sum_{\langle ij \rangle}  c^{\dagger}_i c^{~}_j - 
\mu \sum_{i } n_i  
- \lambda \sum_i n_i  x_i \rangle_{\{x\}} 
~+ ~(K/2)\sum_i x_i^2.  
$
At strong coupling the electronic energy (in angular brackets) 
in a phonon background, $\{x\}$, 
is not analytically known, since the problem is equivalent to
that of electrons in an {\it arbitrary} strongly fluctuating landscape.
The only unbiased way to solve the problem 
is to generate Monte Carlo samples of the phonon configurations and
accept or reject them using the fermion energy obtained by 
exact diagonalisation \cite{ed-ref}.
Such exact diagonalisation based Monte Carlo (ED-MC) 
is feasible only for electronic Hilbert space dimension $H_N$ upto
$\sim 100$ since the computation time $\tau_N$ grows as $ H_N^4$.
Although the finite size effects are quite strong, the small sizes
do provide 
a feel for the phase diagram. 
Our `reference results' in 
2d and 3d are based on ED-MC on $8^2$ and $4^3$   
systems. 

To overcome the severe size limitation of ED-MC we have
used  a `travelling cluster' approximation \cite{tca-ref}
(TCA) for 
estimating the energy change due to a phonon move. Thus, 
instead of estimating the energy change of the update $x_i 
\rightarrow x'_i$, at site ${\bf R}_i$,  as 
$\Delta {\cal E} = {\cal E}\{x'\} - {\cal E}\{x\}$,
by diagonalisation of the full  Hamiltonian, we 
obtain an estimate from a `cluster' Hamiltonian
constructed by considering $L_c^2$ (or $L_c^3$) sites around
the reference site ${\bf R}_i$. The energy of the system as a whole 
of course cannot be approximated by the energy, ${\cal E}_{cl} \{ {\bf R}_i \}$,  
of the  cluster, but 
the {\it change in the energy of the system} when a phonon update is made
is accurately captured by the smaller subsystem.
We have broadly benchmarked TCA against ED-MC in an
 earlier paper
\cite{tca-ref}, and also provide several comparisons between the exact and
TCA results in this paper.
Our large size 
TCA results are obtained by using $4^2$ clusters on $24^2$ systems
in two dimension, and
$4^3$ clusters on $8^3$ in three dimension.

\begin{figure}
\hspace{2cm}
{\epsfxsize=10.0cm, \epsfysize= 9cm, \epsfbox{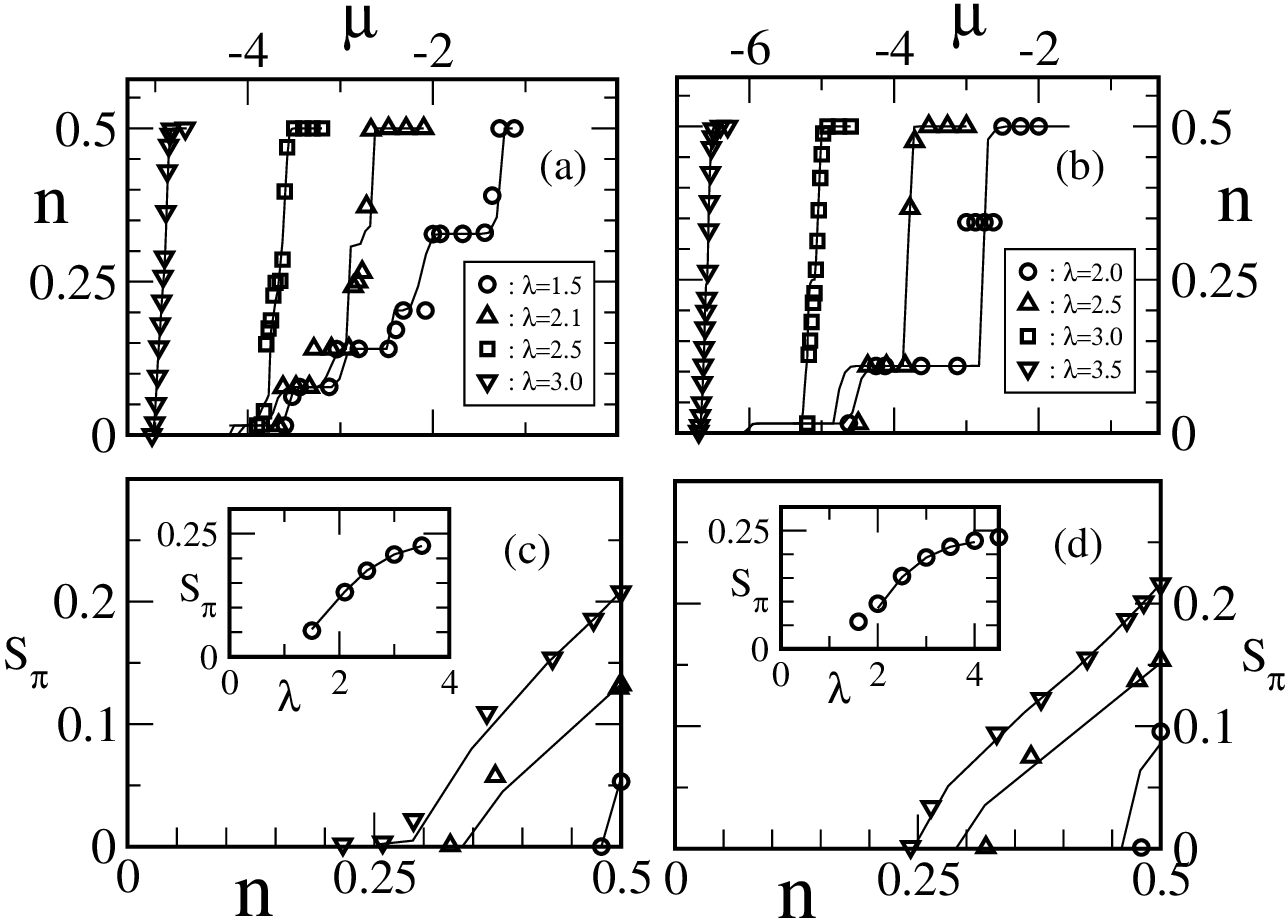}}

\caption{Panels $(a)$ and $(b)$ show the variation in carrier density with $\mu$
computed with ED-MC (symbols) and TCA (lines). Panel $(a)$ is for 2d, with ED
on $8^2$  and TCA with $4^2$ on $8^2$,  while
$(b)$ is for 3d, with ED on $4^3$ and TCA with $3^3$ on $4^3$.
Panels $(c)$~and $(d)$~show the COI `order parameter' with varying electron density and
coupling. Panel $(c)$ is for 2d, panel $(d)$~is for 3d. Symbols for ED, firm lines
for TCA, same system size as in $(a)-(b)$. The insets to the lower panels show
the  COI order parameter at $n=0.5$. 
TCA on large sizes does not significantly 
change the $T\rightarrow 0$ order parameter.
Panel $(a)$ and $(c)$ share  a common legend, 
as  do panel $(b)$ and $(d)$.
}

\vspace{-2mm}

\end{figure}
To allow effective annealing we study the system at low finite temperature,
$T=0.02$,  with varying $\mu$. At equilibrium, at
a given $\mu$, we calculate the following: $(i)$~the electron density, $n(\mu)$, 
$(ii)$~the structure factor 
$S({\bf q})
= (1/N^2) \sum_{ij} \langle \langle  n_i \rangle \langle n_j \rangle \rangle 
e^{i {\bf q}.( {\bf R}_i - {\bf R}_j)}$, including 
$S(  \pi, \pi, ..)$, 
the order parameter for commensurate charge ordering  
($\langle n_i \rangle$ is the quantum average of $n_i$ in a MC 
configuration and the outer angular brackets indicate  average
over configurations),
$(iii)$~the electronic density of states (DOS), $N(\omega)$,
$(iv)$~the distribution of lattice distortions, 
as well as MC snapshots of 
the  electron density.
Fig.1 sets out the ground state phase diagram in 2d and 3d, 
while the detailed indicators,
like $n(\mu)$, $S( \pi, \pi, ..)$, and $N(\omega)$, from which
the phase diagram is constructed, are 
shown in Figs.2-3.
To access the ``weak coupling'' part of the phase diagram, $\lambda < 1.5$ say,
we have used an analytic method described later.

\section{Phases}
The two  panels in Fig.1
show the asymptotic large $L$ phase diagram obtained through a 
combination of TCA and analytic estimates. 
Our simulations  reveal that 
there are primarily three phases in the adiabatic limit, for $T \rightarrow 0$,
in both 2d and 3d. These are $(i)$~the FL phase,
{\it i.e}, the tight binding electron system, without any
lattice distortions, $(ii)$~an insulating polaron liquid 
(PL) with `liquid like' 
short range  positional correlations, and a gap 
in the DOS,
 and $(iii)$~charge ordered insulating (COI) 
phases, with a gap in the DOS, and a peak at $ {\bf q} = \{\pi, \pi,..\}$
in $S({\bf q})$. The COI is either a simple ``chessboard'' phase at $n=0.5$, 
or a phase with ``defects'' 
off $n=0.5$ (and strong coupling).
There is also the possibility of other charge ordered phases,
apart from $\{ \pi, \pi, ..\}$, in the vicinity of $n=1/4, 1/8$,
{\it etc}, at strong coupling, but their 
energy difference with respect
to the PL phase is very small, making it difficult to access them in a
low but finite $T$ simulation.
There are no ``metallic'' phases with lattice distortions, in contrast
to the DMFT results in the adiabatic limit \cite{millis-mull}.
\begin{figure}
\hspace{2.0cm}
{\epsfxsize=11.0cm, \epsfysize= 10cm, \epsfbox{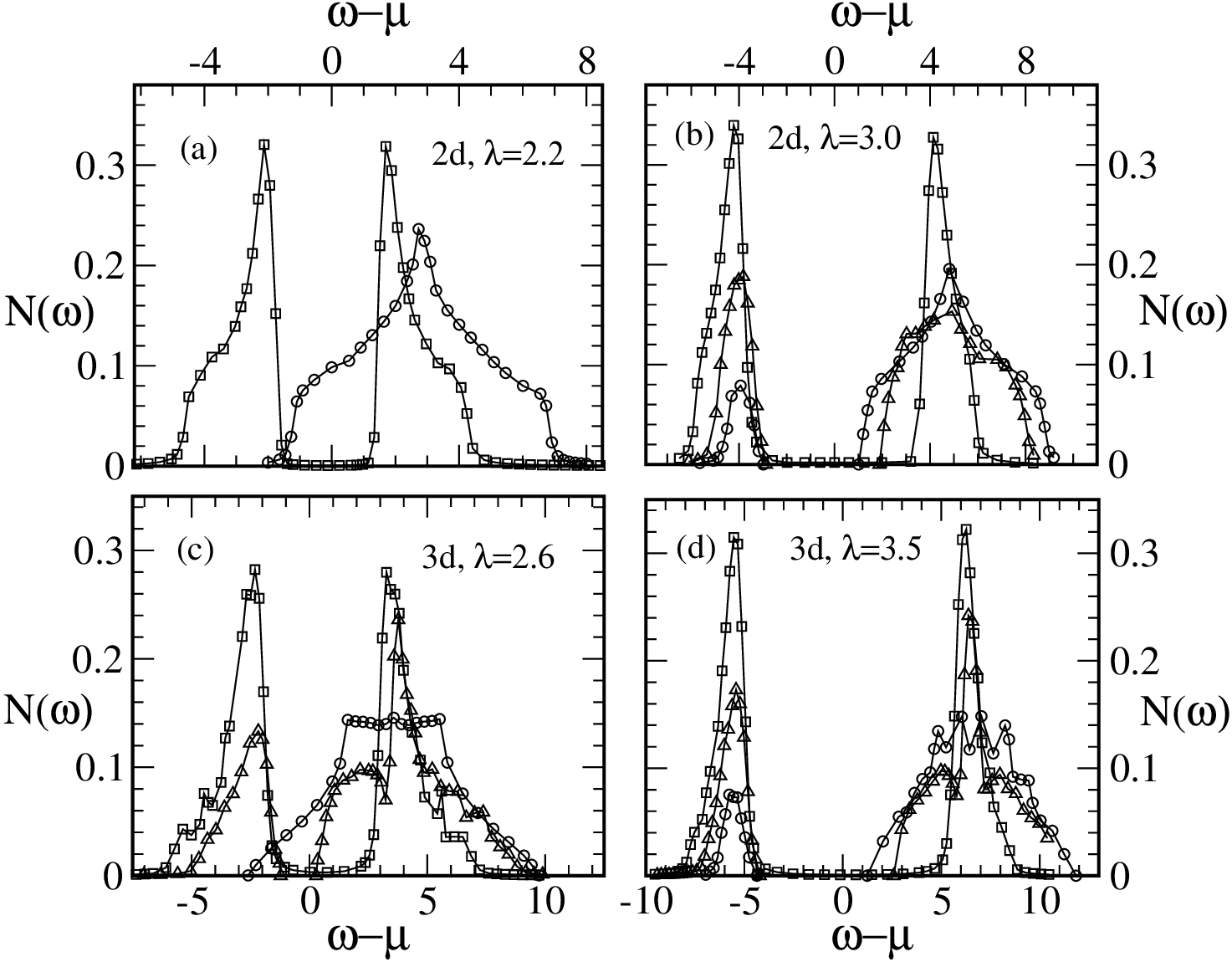}}

\caption{Density of states computed with TCA on $24 \times 24$ in 2d, and
$8 \times 8 \times 8$ in 3d. Symbols: circles for $n=0.1$,
triangles for $n=0.3$, and squares for $n=0.5$.  $(a)$ 2d, $\lambda = 2.2$, 
$(b)$~2d, $\lambda= 3.0$,
$(c)$~3d, $\lambda = 2.6$, $(d)$~3d, $\lambda = 3.5$.
}

\vspace{-2mm}

\end{figure}
We discuss the phase diagram in terms of the two broad regimes:
$(i)$~strong coupling polaronic phases, and $(ii)$~weak coupling: the
 COI instability and phase separation.

{\it $(i)$~Strong Coupling:} A naive balance 
of the `polaron binding energy'
$E_p = \lambda^2/2K$ (for a site localised electron), 
and the lower edge of the tight binding band, $-E_b = 2dt$, suggests
that polaron formation would occur at 
$\lambda_c  \approx \sqrt{4Kdt}$, implying  
$\lambda_c^{2d} \sim 2.82$ and $\lambda_c^{3d} \sim 
3.46$. More accurate earlier  estimates \cite{pol-ref2}, 
and our simulations,
indicate $\lambda_c^{2d} \sim 2.6$ and $\lambda_c^{3d} \sim 3.3$.
The lowering arises because the polaron is not quite site
localised, and taking into account the kinetic energy $\sim dt^2/E_p$,
the polaronic state becomes favourable slightly before the naive
threshold.  
The polaron states  which emerge for $\lambda \ge \lambda_c$ are
compact, although not site localised, 
with wavefunctions decaying over a couple of lattice spacings. 
Due to the strong localisation, and the associated short range
lattice distortion field, the {\it finite density}
system is  a  short range correlated array of trapped
electrons (the PL phase), keeping maximum mutual separation in order 
to maximise kinetic energy gain 
by virtual hopping. 
In the  dilute limit the polaron system has a very large compressibility,
$\partial n/{\partial \mu}$, since the
polarons are effectively non interacting and $\mu$ is virtually
`pinned' to the single polaron energy as the carrier density is increased
from zero. This dictates the sharp rise in the $n-\mu$ curves, Fig.2.$(a)-(b)$.
However, with $n$ increasing towards half-filling the short range `repulsion',
$\sim {\cal O}(t^2/E_p)$ for nearest neighbours, 
begins to be felt and the $n-\mu$ characteristic is flat at $n=0.5$,
indicative of an incompressible half filled state.
The polaronic 
repulsion favours a ``chessboard'' CO phase at $n=0.5$.

The off half-filling phase maintains the positional correlations of the $n=0.5$
phase, but has vacancies (or defects), see {\it e.g}, Fig.4. The polarons tend 
to avoid nearest neighbour locations, preferring to be along the diagonal (where 
the `repulsion' is weaker). As long as the system is sufficiently dense, these
positional correlations `percolate' sustaining 
long range order, with a clear peak in
$ S( \pi, \pi, ..)$. Our results suggest that the strong coupling 
off half-filling COI phase
survives down to a density $\approx 0.35$
in the coupling regime shown (Fig.1). 
The order parameter for the COI phases 
is shown in Fig.2.$(c)-(d)$, and the DOS in
Fig.3. The order parameter computed with TCA on small sizes, 
see Fig.2 caption, accurately matches
ED-MC results. TCA results on large sizes for $S( \pi, \pi, ..)$ 
(not shown)
are not significantly different. 

The critical coupling for the FL to PL transition 
is {\it density dependent},
decreasing significantly, Fig.1~$(c)-(d)$,
from the single polaron threshold as $n$ is 
increased. By the time 
$n \sim 0.3-0.4$, $\lambda_c$ is reduced to $\approx 75 \%$ of
the `single polaron' value, in both 2d and 3d, {\it i.e} the critical
$E_p$ is almost halved.
Qualitatively, electrons which are already  localised 
affect an added particle via their strongly 
inhomogeneous distortion field, $\{ x\}$, which  interplays with EP
coupling leading self consistently to the decreasing $\lambda_c(n)$. 
Trapping at finite $n$ involves a combination of 
(self generated) ``disorder''
and polaronic tendency \cite{emin-buss}, akin to the interplay of Anderson
localisation and single polaron formation.
All the phases, COI or PL,  which occur for $\lambda \ge \lambda_c(n)$ have a 
clear spectral gap, and are distinguished mainly by $S({\bf q})$.

\begin{figure}
\hspace{2.0cm}
{\epsfxsize=10cm, \epsfysize= 7cm, \epsfbox{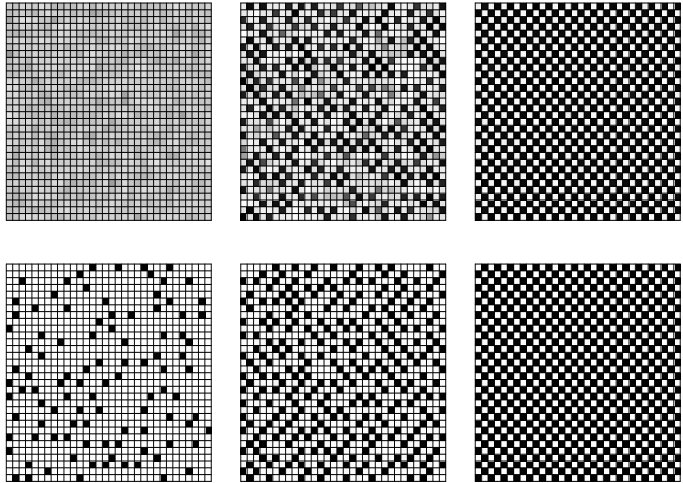}}

\caption{Monte Carlo snapshots of the electron density $n_{\bf r}$ with
varying $n$ and $\lambda$ in 2d. Top panels $\lambda = 2.2$, bottom 
panels $\lambda = 3.0$. Density $n = 0.10, 0.30, 0.50$, left to right.
Top left panel is FL. Both panels to the extreme right are `perfect'
COI, while the other three panels are PL.}

\end{figure}

{\it $(ii)$~Weak Coupling:}
At weak coupling there is an instability in a bipartite tight binding
lattice, at $n=0.5$, due to Fermi surface nesting. The density response
function $\chi({\bf q}, \omega=0)$ diverges for ${\bf q} 
\rightarrow \{ \pi, \pi, ..\}$. For $\lambda \rightarrow 0$, 
where the analysis in terms of the non interacting susceptibility is
valid, this leads to a charge order instability, with ordering temperature 
$T_{CO} \sim  te^{-t/E_p}$. 
With increasing $\lambda$ the CO phase continues to be the ground state,
but as  the charge modulation increases, and the
band splitting grows, the FL to COI transition at $T=0$ becomes
{\it first order}, with respect $n$,
since the electron density in the modulated $x_i$ 
background is significantly different from that in the homogeneous background.
The $n-\mu$ data in Fig.2.~$(a)-(b)$ highlight the discontinuity in $n$
near half-filling.
It is difficult to access the instability and the coexistence
width at weak coupling  in a small system since 
$\chi({\bf q}, \omega=0)$ has strong size dependence. Direct simulation can 
locate the CO phase only down to $\lambda \sim 1.5$.  
At lower $\lambda$ we variationally 
computed the COI order parameter at $n=0.5$ on large lattices, 
 and also the $\mu_c(\lambda)$ at which
the FL to COI transition occurs. It suggests that 
the FL to COI transition becomes (noticeably) 
first order, with a density discontinuity,
for $\lambda \ge 1.0$, in both 2d
and 3d.
The  weak coupling COI phase at $n=0.5$
connects continuously to the strong coupling ``polaron ordered'' phase
\cite{ciuchi-cdw,millis-cdw}.

\section{Discussion}
Despite accurate handling of strong coupling and spatial correlations, a few 
physical effects in the adiabatic limit are still difficult to capture within 
a numerical simulation. Other effects, of possible relevance to real materials,
require us to go beyond the adiabatic limit itself. This section briefly discusses
these issues.
$(i)$~As we have discussed, even in the adiabatic limit, there
are possibly additional commensurate CO phases at $n= 1/4, 1/8$ etc, which could
show up at very low temperature. There could also be {\it incommensurate} CO
phases at weak coupling, these are hard to access on small systems,
given the strong finite size effects in $\chi({\bf q}, \omega=0)$.
$(ii)$~In real materials $\omega_{ph}$ would be finite, and quantum
fluctuations  would modify some of our conclusions:
$(a)$~The simple `band metal' (FL) obtained  at $\omega_{ph}=0$ 
would  become a  correlated Fermi liquid, with effective 
mass renormalisation, band narrowing, {\it etc},
if $\omega_{ph} \neq 0$.
$(b)$~Quantum fluctuations would restore translation invariance in the PL
phase \cite{ciuch-prl} so that below a coherence scale, $T_{coh} \sim \omega_{ph}$, 
the resistivity falls quickly to zero, instead of diverging as the gapped DOS here 
would suggest.  
$(c)$~The COI phase would survive  for $n \sim 0.5$ 
but possibly with reduced ordering temperature.
Overall,  even with finite $\omega_{ph}$, 
as long as $\omega_{ph}/t \ll 1$, our ``ground state'' phase diagram would be
qualitatively useful for  $T >  T_{coh}$  in the model.

In summary, we have used a new real space method  that naturally incorporates
the spatial correlations vital at metallic densities, to clarify the phase
diagram of the adiabatic Holstein model.
This approach allows  unconstrained optimisation, 
not restricted to any specific kind of order, and  readily reveals the 
regimes of phase coexistence. 
It also extends naturally 
to incorporate the effects of disorder and thermal fluctuations \cite{sk-pm-latt-pol2}.
Another possibility, we will separately report,
is to include spin degrees of freedom, via double exchange, 
to model the phenomena in manganite physics.

\vspace{.1cm}

{\it Acknowledgement} We acknowledge use of the Beowulf cluster at H.R.I.

{}

\end{document}